\documentclass[aps,prl,reprint,nofootinbib,superscriptaddress]{revtex4-1}
\usepackage[utf8]{inputenc}
\usepackage{amsmath,amssymb,bbm,color,lmodern,graphicx}
\usepackage[section]{placeins}
\begin{document}
\title{Robust parameter determination in epidemic models with analytical descriptions of uncertainties}  
\author{G.M. Nakamura}
\affiliation{Faculadade de Filosofia, Ciências e Letras de Ribeirão
  Preto (FFCLRP) \\
Universidade de S\~{a}o Paulo (USP), 14040-901 Ribeir\~{a}o Preto,
Brazil}
\affiliation{Instituto Nacional de Ci\^{e}ncia e Tecnologia -
  Sistemas Complexos (INCT-SC)}
\author{N.D. Gomes}
\affiliation{Faculadade de Filosofia, Ciências e Letras de Ribeirão
  Preto (FFCLRP) \\
Universidade de S\~{a}o Paulo (USP), 14040-901 Ribeir\~{a}o Preto,
Brazil}
\author{G.C. Cardoso}
\affiliation{Faculadade de Filosofia, Ciências e Letras de Ribeirão
  Preto (FFCLRP) \\
Universidade de S\~{a}o Paulo (USP), 14040-901 Ribeir\~{a}o Preto,
Brazil}
\author{A.S. Martinez}
\affiliation{Faculadade de Filosofia, Ciências e Letras de Ribeirão
  Preto (FFCLRP) \\
Universidade de S\~{a}o Paulo (USP), 14040-901 Ribeir\~{a}o Preto,
Brazil}
\affiliation{Instituto Nacional de Ci\^{e}ncia e Tecnologia -
  Sistemas Complexos (INCT-SC)}
\begin{abstract}   
  Compartmental equations are primary tools in disease spreading
  studies. Their predictions are accurate for large populations but
  disagree with empirical and simulated data for finite populations,
  where uncertainties become a relevant factor. Starting from the agent-based
  approach, we investigate the role of uncertainties and autocorrelation
  functions in SIS epidemic model, including their relationship with
  epidemiological variables. We find new differential equations
  that take uncertainties into account. The findings provide
  improved predictions to the SIS model and it can offer new insights
  for  emerging diseases. 
\end{abstract}
\maketitle

Communicable diseases are health disorders caused by pathogens
transmitted from infected individuals to susceptible ones 
\cite{bonita2006}. In general, the transmission process occurs with
variable success rate, subjected to stochastic uncertainties during
the infectious period of the host. These uncertainties comprehend
aspects related to transmission mechanisms and availability of adequate
contact between hosts and susceptible individuals. The latter aspect
has been further improved  {via} network
theory, accounting for more realistic social interactions and 
highlighting the role of central hubs in general disease spreading
dynamics \cite{albertRevModPhys2002, satorrasRevModPhys2015,
  keelingPNAS2002, keelingJRSoc2005, bansalJRSoc2007}. 
For large and well-connected populations, stochastic factors are
discarded in favor of differential equations, also known as
compartmental equations \cite{kermackProcRSocA1927}.   
Generalizations for compartmental equations have been able to
reproduce pandemics and extract relevant characteristics, taking into
account more complex network topologies \cite{satorrasRevModPhys2015,
  satorrasPhysRevLett2001, satorrasPhysRevE2001, brockmannNJPhys2017}.

In contrast, the stochastic nature of disease transmission cannot be
omitted for a number of scenarios. It becomes more pronounced for
small populations. In this case, the individual characteristics of each
agent forming the population are relevant variables to the spreading 
process. Incidentally, this is often the case of emerging diseases
\cite{heesterbeekScience2015}. Because the population cannot be
treated as homogeneous, average values for the population are no
longer adequate and the accuracy of compartmental equations decreases
for increasing uncertainties. 
Stochastic models deal with this issue by proposing simpler rules to
express the disease transmission, taking the relevant stochastic
factors into account. Besides average values, stochastic models
possess additional tools to provide further insights, including
autocorrelation functions. For instance, in the standard Brownian
motion,  
the delta-like behavior observed for the white-noise autocorrelation
function dictates the linear dependence between spatial variance and
time. In disease spreading processes, however, autocorrelation
functions have been largely neglected.

Here, we study the role of uncertainties and the normalized autocorrelation
function, $D_{\rho\rho}(t)$, in the SIS epidemic model
for a population with $N$ agents.
From $D_{\rho  \rho}(t)$, we derive the differential equation that
governs the dynamics of the variance $\sigma^2(t)$ associated with the
average density of infected agents $\langle \rho(t) \rangle$ in the
population. We build a system of differential equations
to describe the SIS model and validate them with numerical
simulations. In addition, we briefly discuss the manner in which the
Fano factor affects the extraction of epidemiological parameters.



\textit{Compartmental models.}
Let $\rho(t)$ be the density of infected agents in a population of
size $N$ in the SIS model. In the compartmental approach, the
population is assumed to be large, homogeneous and highly
interconnected. As a result, agents can be regarded as
statistically equivalent. This implicit assumption is equivalent to
complete the permutation symmetry, which is also found in the complete
graph \cite{albertRevModPhys2002}.  
Thus, $\rho(t)$ becomes the key variable in the compartmental approach.  

The other relevant assumption concerns the transmission
mechanism. Because the population is taken as homogeneous, the
adequate interaction between infected and susceptible agents
occurs with probability proportional to  $(1-\rho) \rho$. This
assumption constitutes the basis for the random mixing hypothesis 
\cite{keelingJRSoc2005}.  At the same time, recovery events are
proportional to the infected density $\rho$. Adding both
contributions, the SIS compartmental equation for infected density is  
written as 
\begin{equation}
  \label{eq:sis}
  \frac{d \rho}{d t} = \alpha \left( 1 - \rho \right) \rho - \gamma \rho,
\end{equation}
where $\alpha$ and $\gamma$ are the transmission and recovery
rate, respectively. A well-accepted generalization proposed in
Ref.~\cite{satorrasPhysRevLett2001} takes network metrics into account
in the transmission rate, improving the overall accuracy of
Eq.~(\ref{eq:sis}) for complex networks \cite{satorrasRevModPhys2015}.

Rearranging Eq.~(\ref{eq:sis}), we obtain
\begin{equation}
  \label{eq:sis2} 
  \frac{d}{d t}\ln \rho = \alpha(\rho_{\textrm{eq}} -  \rho),
\end{equation}
where $\rho_{\textrm{eq}}=1-\gamma/\alpha$ is the steady state density
for $\alpha > \gamma$. 
From 
Eq.~(\ref{eq:sis2}), one can extract $\alpha$ and $\gamma$ by  a
linear fit. 
Furthermore, using the actual solution of
Eq.~(\ref{eq:sis}) in Eq.~(\ref{eq:sis2}) leads to:
\begin{equation}
  \label{eq:sis3}
  \frac{1}{\rho}\frac{d}{d t}\ln \rho =
  {\alpha}\left[ \frac{\rho_{\textrm{eq}}}{\rho(0)} -  1 \right]
  \textrm{e}^{-\alpha\rho_{\textrm{eq}}t}, 
\end{equation}
whose decay rate depends only on epidemiological
parameters.

\begin{figure}[htb]
  \includegraphics[width=0.95\columnwidth]{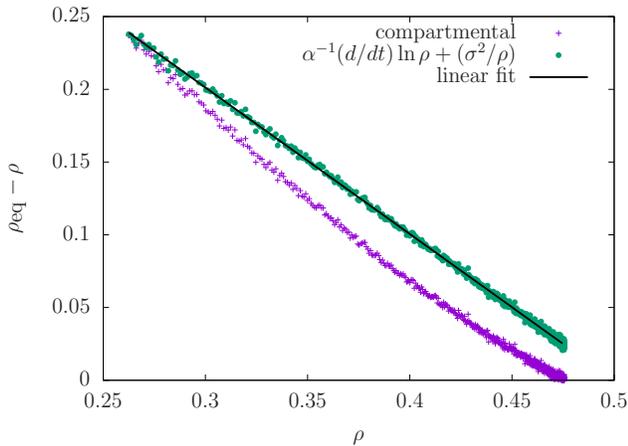}
  \caption{\label{fig:compartmental} Deviations of compartmental predictions. Values of
    $\rho_{\textrm{eq}}-\rho$ versus $\rho$ using the formula from
    compartmental equation $(d/dt)\ln \rho$ without corrections
    (cross), with corrections that depend on Fano factor
    $\sigma^2/\rho$ (full circle). MC simulations are performed
    using $10^6$ samples for complete graph with $N=50$ agents,
    $\gamma = 1/2$ and $\alpha = 1$. Linear fit (solid line) produces
    $\gamma_{\textrm{data}}=0.50(3)$ and $\alpha_{\textrm{data}}=1.00(0)$.}
\end{figure}

It should be clear by now that Eq.~(\ref{eq:sis}) is an important tool 
to extract epidemiological parameters. What would be the implications
for epidemiological studies if Eq.~(\ref{eq:sis}) had additional terms
or corrections? The current methodology to evaluate $\alpha$ and
$\gamma$ would carry systematic errors. Fig.~(\ref{fig:compartmental}) 
displays the values of $\alpha^{-1}(d/dt)\ln \rho$ using
Eq.~(\ref{eq:sis2}) obtained from numerical simulations. It deviates from
$\rho_{\textrm{eq}}-\rho$. Even more, parameter estimation for $\rho$
that is typical during the onset of epidemics \textit{underestimates} the
transmission rate. If the Fano factor is taken into account, however,
we recover the linear behavior, as we discuss in what follows.


\textit{Agent-based models.}
In the stochastic approach, the population consists of $N$
distinguishable agents connected to each other according to a
pre-defined adjacency matrix $A$ ($N\times N$). In the complete graph,
each agent interacts with the remaining $N-1$ agents, $A_{i j} = 1
-\delta_{i j}$. Each agent ($k=0,1,\ldots,N-1$) may assume one of two
possible health states $n_k$ in the SIS model, either susceptible
($n_k=0$) or infected ($n_k=1$). Following
Ref.~\cite{nakamuraSciRep2017}, there are $2^N$ available
configurations in the canonical basis $\lvert \mu \rangle$, with $\mu=
0,1,\ldots,2^{N-1}$. Configurations are readily extracted
from the binary construction $\mu = n_02^0+n_1 2^1+\cdots+
n_{N-1}2^{N-1}$. As an example, for $N=4$, the configuration $\lvert 0
\rangle = \lvert 0 0 0 0\rangle$ represents the infected-free
configuration, whereas all agents are infected in $\lvert 15\rangle =
\lvert 1 1 1 1\rangle$.

In this paper, we treat the disease spreading process as a Markov
process. Following Ref.~\cite{nakamuraSciRep2017}, the master
equation in operator notation
is
\begin{equation}
  \frac{d }{d t} \lvert P(t)\rangle = -\hat{H} \lvert P(t)\rangle,
  \label{eq:fokker}
\end{equation}
in which $\lvert P(t)\rangle= \sum_{\mu=0}^{2^N-1}P_{\mu}(t) \lvert \mu
\rangle $ is the probability vector, with $P_{\mu}(t)$ being the
instantaneous probability to find the system in the configuration
$\lvert \mu \rangle$; and $\hat{H}$ is the generator of time
translations, given by the following expression:
\begin{equation}
  \hat{H} = \frac{\alpha}{N}\sum_{k,\ell = 0}^{N-1}A_{k\ell}(1 -
  \hat{n}_k-\hat{\sigma}_k^+)\hat{n}_{\ell} +\gamma \sum_{k=0}^{N-1}(\hat{n}_k-\hat{\sigma}^-_k).
  \label{eq:hami}
\end{equation}
The operators $\hat{n}_k$ extract the health state of the $k$-th
agent, $\hat{n}_k\lvert n_0\cdots n_k\cdots \rangle = n_k \lvert
n_0\cdots n_k\cdots \rangle$, while $\hat{\sigma}_k^{\pm}$ are the usual
spin-$1/2$ ladder operators. Operators are assigned the hat symbol to
distinguish them from scalars.

\begin{figure*}
  \includegraphics[width=0.42\textwidth]{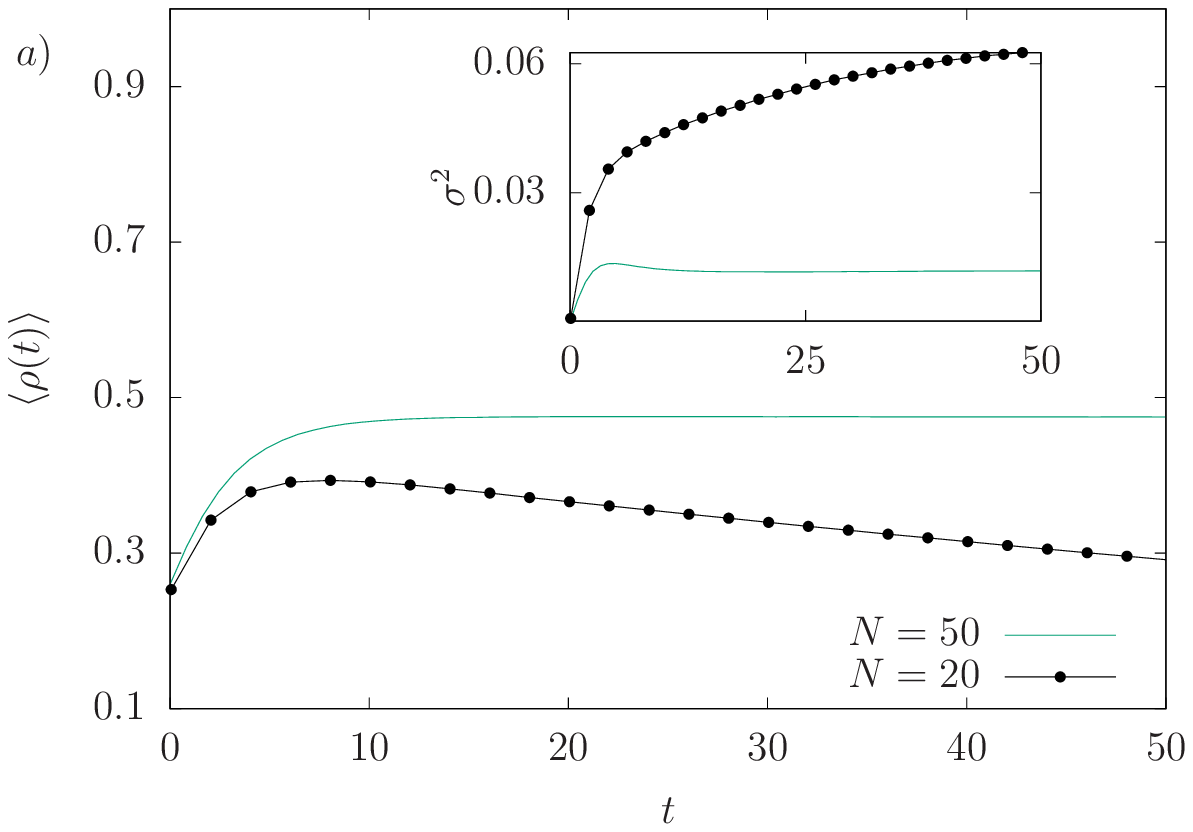}~
  \includegraphics[width=0.42\textwidth]{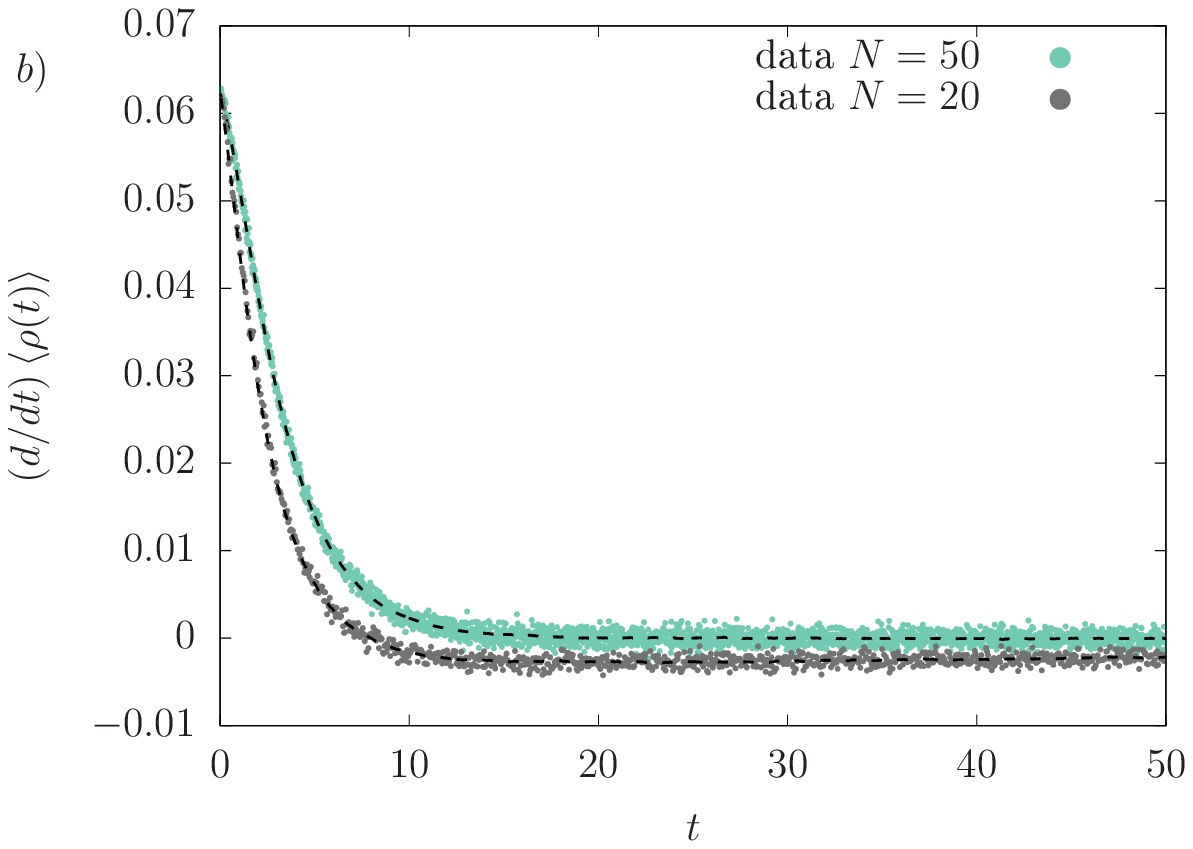}\\
  \includegraphics[width=0.42\textwidth]{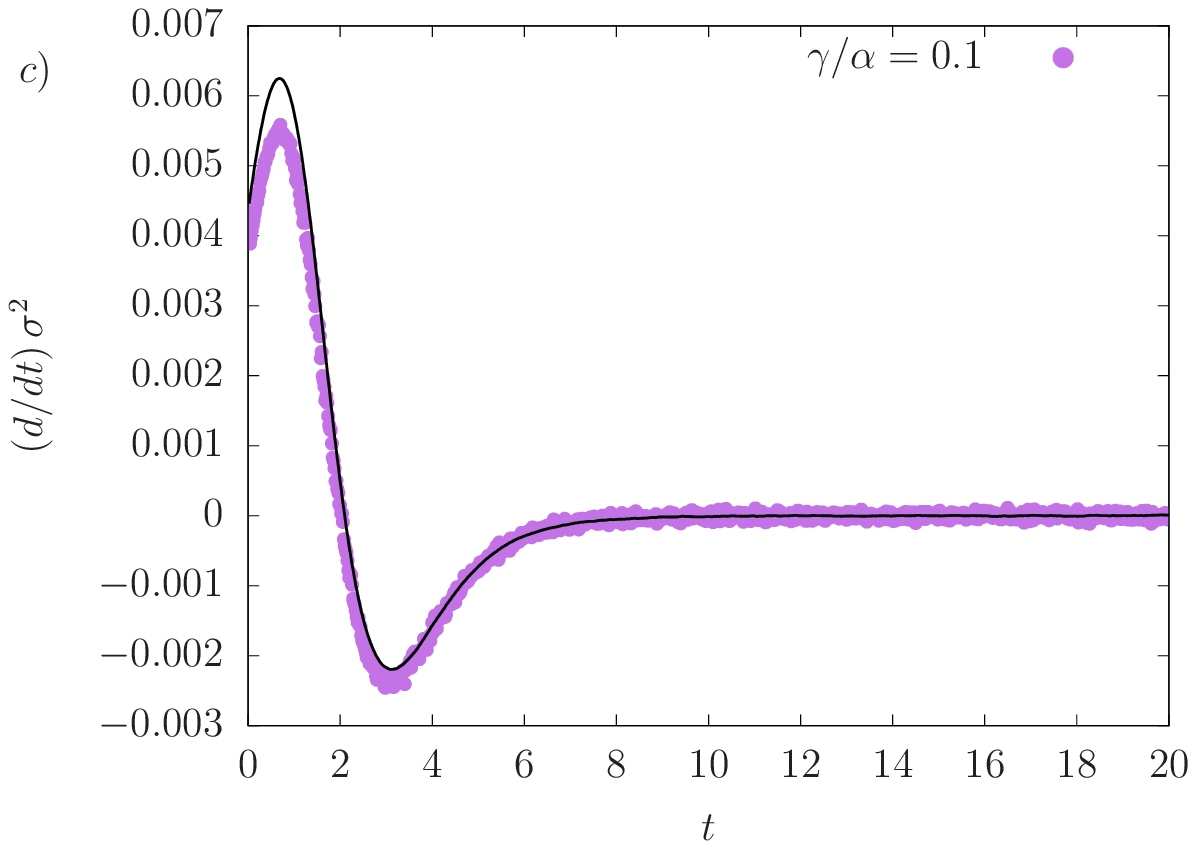}~
  \includegraphics[width=0.42\textwidth]{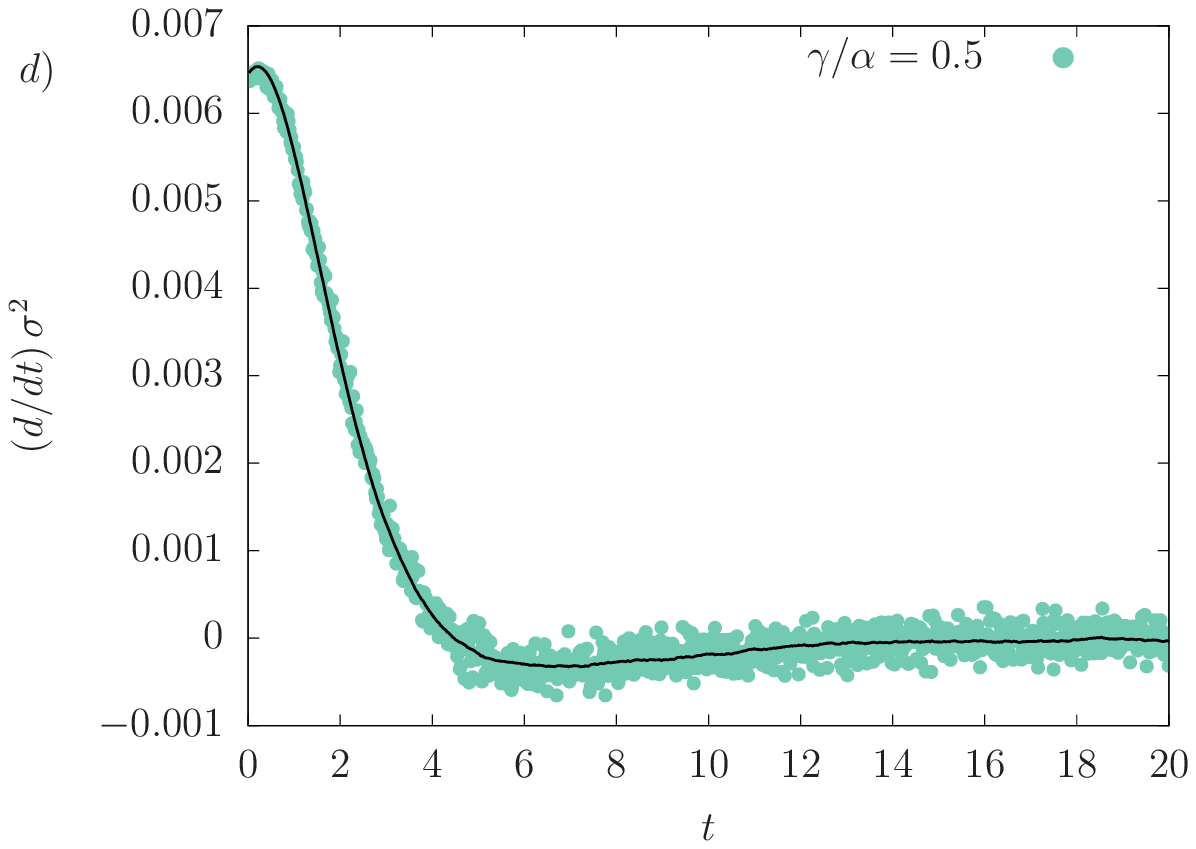}
  \caption{\label{fig:complete_density} Uncertainty in agent-based
    simulations. a) Density of infected agents for  $10^6$
    simulations, with $\gamma/\alpha = 1/2$, and $N=20$ (full 
    circle) or $N=50$ (solid line) agents. The curve with $N=50$ is
    compatible with compartmental equations, while $\langle
    \rho(t)\rangle $ decays for the case with $N=20$ agents. (Inset)
    Uncertainties become more prominent in smaller populations, suggesting
    finite size corrections. b) Temporal derivative of $\langle \rho
    \rangle$ using formula in Eq.~(\ref{eq:del-rho}) (dashed lines) and
    their corresponding values calculated directly from data (circles).
  c) Forward time derivative of $\sigma^2(t)$ for $N=50$ agents and
  $\gamma/\alpha=0.1$ (circles). The solid line represents the values
  extracted using Eq.~(\ref{eq:del-sigma}). Same as c) but
  $\gamma/\alpha=1/2$. }  
\end{figure*}

Eqs.~(\ref{eq:fokker}) and (\ref{eq:hami}) can be
used to evaluate 
the average density of infected agents,
\begin{equation}
  \langle \rho(t) \rangle \equiv
  \frac{1}{N}\sum\limits_{\mu=0}^{2^N-1} \eta_{\mu} P_{\mu}(t),
\end{equation}
where $\eta_{\mu}\equiv  \sum_{k}\langle \mu \vert \hat{n}_k \vert
\mu\rangle$ is the number of infected agents in the configuration
$\lvert\mu\rangle$. 
Exploiting the fact that $\sum_{\mu}\eta_{\mu} \langle \mu \vert
\sum_{k}\hat{\sigma}^+_k \vert \nu \rangle =
(\eta_{\nu}+1)(N-\eta_{\nu})$ and $\sum_{\mu} \eta_{\mu}\langle 
  \mu\vert \sum_k \hat{\sigma}_k^-\vert \nu \rangle =
  (\eta_{\nu}-1)\eta_{\nu}$, in the complete graph,
\begin{equation}
  \frac{d\langle \rho \rangle}{d t} =  \alpha\left[\rho_{\textrm{eq}} -\langle
  \rho(t)\rangle\right]\langle \rho(t)\rangle - \alpha \sigma^2(t),
  \label{eq:del-rho}
\end{equation}
with instantaneous variance $\sigma^2(t) = \langle \rho^2\rangle -
\langle \rho\rangle^2$. Eq.~(\ref{eq:del-rho}) exhibits excellent
agreement with numerical simulations (see
Fig.~{\ref{fig:compartmental}}), and recovers the compartmental equation
Eq.~(\ref{eq:sis}) for vanishing $\sigma^2(t)$. Fig.~\ref{fig:complete_density}
depicts $\langle \rho(t)\rangle$ in the complete graph for
$N=20$ and $50$ agents. 
The case with $N=20$ deviates from compartmental results:
fluctuations that eradicate the disease are more likely to occur in
scenarios with small populations, even if agents are statistically
equivalent. 

We emphasize that the inherent fluctuations of the disease spreading
process is summarized by $\sigma^2(t)$ in Eq.~(\ref{eq:del-rho}). An 
initial uncertainty evolves during the time evolution of
$\rho(t)$, reinforced by the fact that agents can only be either
susceptible or infected, {i.e.}, they obey Fermi-Dirac
statistics. In a sense, $\sigma^2(t)$ is conceptually similar to the
shot noise in condensed matter physics \cite{blanterPhysRep2000}.

As the plots of Fig.~\ref{fig:complete_density} show, numerical simulations support
Eq.~(\ref{eq:del-rho}) predictions with good accuracy, highlighting 
the role of uncertainties in the SIS model. Noting that $\sigma^2(t)$
depends on time, there must exist an additional differential 
equation for $\sigma^2(t)$  (see
Fig.~\ref{fig:complete_density}). Indeed, the same rationale behind Eq.~(\ref{eq:del-rho}) can be used
to find $(d/dt)\sigma^2$: 
\begin{align}
  \frac{1}{2\alpha}\frac{d\sigma^2}{dt}
  &= \left[ \langle \rho\rangle+
  \rho_{\textrm{eq}}  -\frac{1}{N}
    \right]\sigma^2(t) - \left[ \langle \rho^3\rangle - \langle
    \rho\rangle^3\right] +
    \nonumber\\
  \label{eq:del-sigma}
  &-     \left[ \frac{\langle   \rho\rangle +
    \rho_{\textrm{eq}}}{2} - 1\right]
    \frac{\langle
    \rho\rangle}{N}.
\end{align}
Even if $o(1/N)$ corrections are omitted, one still must take into
account the contributions from  $\Delta_3(t) \equiv \langle
\rho^3\rangle - \langle \rho\rangle^3$. Formally, we could calculate the
differential equation for $\Delta_3(t)$ but then we would have to deal
with $\langle \rho^4(t)\rangle$ and so on.

Let us briefly assume that is possible to estimate $\Delta_3(t)$
without higher statistical moments. In this case, 
Eqs.~(\ref{eq:del-rho}) and (\ref{eq:del-sigma}) form a system 
of differential equations for $\rho(t)$ and $\sigma^2(t)$. Therefore,
our main task is to obtain surrogate dynamics for $\Delta_3(t)$, which
likely depend on the behavior or nature of the fluctuation itself. In
fact, the density autocorrelation function provides valuable insights
on $\Delta_3(t)$ for non-symmetric fluctuations. Likewise, the
existing relationship between $\Delta_3(t)$ and the instantaneous
coefficient of skewness, $\kappa_3(t)$, provides a way to investigate
symmetric fluctuations.



\begin{figure}
  \includegraphics[width=0.95\columnwidth]{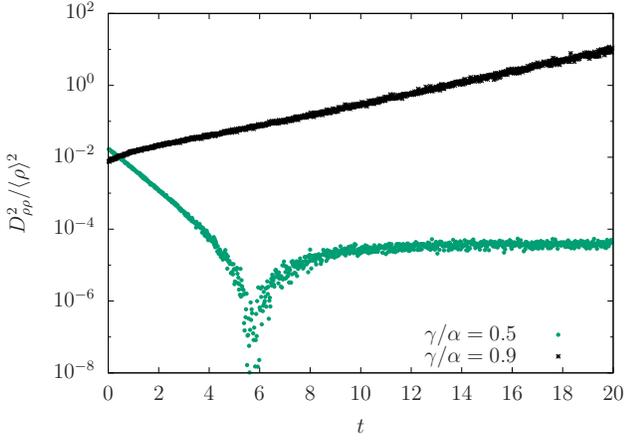}
  \caption{\label{fig:complete_normalized_corr} Contributions
    for $|D_{\rho \rho}(t)/\langle \rho\rangle|^2$. Simulations results
    comprehend $10^6$ simulation samples in the complete graph with
    $N=50$. Gaussian fluctuations are present for
    $\gamma/\alpha=0.5$ (green circles). An exponential decay is observed during the
    outbreak onset. The divergence appears as $\langle
    \rho\rangle$ approaches  $\rho_{\textrm{eq}}$. Finite size
    corrections drive $\langle \rho(\infty)\rangle$ to slightly lower
    values than  $\rho_{\textrm{eq}}$ in the steady state.  Non-Gaussian
    fluctuations and finite size effects create exponential growth for
    $\gamma/\alpha=0.9$ (black cross). }  
\end{figure}

\textit{Autocorrelation function.} For typical disease spreading
processes, the correlations between the various agents that comprise the
finite system are usually weak. So it might seem counterintuitive to assume
that correlations are relevant statistics in epidemic models. However,
autocorrelation functions and variances share similar magnitudes.
Therefore, there is no ground to discard one and keep the other unless
proven otherwise.

Let $C_{\rho \rho}(t)$ be the instantaneous autocorrelation function
between $\rho(t)$ and $\rho(t+\delta t)$, lagged by a single time
window: 
\begin{equation}
  \label{eq:corr}
  C_{\rho \rho}(t) \equiv \langle \rho(t+\delta t) \rho(t)\rangle - \langle
  \rho(t) \rangle^2. 
\end{equation}
Here, averages are evaluated by considering samples from an ensemble
instead of usual Fourier transform, as the ergodic hypothesis is
unavailable. For Markov processes,
\begin{equation}
  \langle
\rho(t+\delta t) \rho(t) \rangle = \frac{1}{N^2}\sum_{\mu}\sum_{k,j}\langle
\mu\vert  \hat{n}_k 
\textrm{e}^{-\hat{H} \delta t}  \hat{n}_j \vert P(t)\rangle.
\end{equation}
The evaluation of this expression involves the same rationale used
for Eq.~(\ref{eq:del-rho}). Plugging the result into
Eq.~(\ref{eq:corr}) we find the relation between $\langle \rho^3(t)\rangle$
and $C_{\rho\rho}(t)$, namely, $ C_{\rho \rho}(t) - \sigma^2(t)  =
\alpha\delta t\left[\rho_{\textrm{eq}}  (\langle
  \rho\rangle^2+\sigma^2) - \langle  \rho^3\rangle\right]   +o(\delta 
t^2)$. Unfortunately, $C_{\rho \rho}(t)$ does not exhibit a simple
functional form.

Instead, consider the normalized autocorrelation function:  
\begin{equation}
  \label{eq:normalized}
  D_{\rho\rho}(t)\equiv 
  \alpha\left[\rho_{\textrm{eq}} -\frac{\langle \rho^3\rangle}{\langle
      \rho\rangle^2}\right] + \alpha \rho_{\textrm{eq}}\frac{\sigma^2}{\langle \rho\rangle^2}.
\end{equation}
Note that $D_{\rho\rho}(t)$ recovers Eq.~(\ref{eq:sis2}) when
$\sigma^2 \rightarrow 0$ and $N\rightarrow\infty$:
$D_{\rho\rho}(t) \rightarrow
\alpha[\rho_{\textrm{eq}}-\langle\rho(t)\rangle] $. Hence $D_{\rho\rho}(t)$ can be 
interpreted as an alternative metric to describe the evolution of  
the system. We can explore this interpretation to learn more about  
$D_{\rho\rho}(t)$ and its time dependence. For instance, using
Eq.~(\ref{eq:sis3}) as inspiration, $D_{\rho\rho}(t)/\langle \rho(t)\rangle$ exhibits
exponential behavior during transient regimes (see
Fig.~\ref{fig:complete_normalized_corr}) regardless of fluctuation
type.
Thus, $D_{\rho\rho}(t)$ is a suitable quantity to express
$\Delta_3(t)$ in Eq.~(\ref{eq:del-sigma}):
\begin{equation}
  \frac{1}{2\alpha}\frac{d}{dt}\ln\sigma^2 = \langle\rho \rangle
  +\left[\langle\rho \rangle - {\rho_{\textrm{eq}}} + \frac{1}{\alpha}{D_{\rho\rho}}
    \right]\frac{ \langle\rho \rangle^2}{\sigma^2}
  +\frac{s}{N},
\end{equation}
with finite size contributions 
$s(t)\equiv s = (\rho/2\sigma^2)(2 - \rho-\rho_{\textrm{eq}})  -1$.

We can gain further insights about $D_{\rho\rho}(t)$ in the case in
which the variance remains finite but $\langle \rho(t)\rangle$ decays
exponentially with decay rate $\tau$. Because $\sigma^2(t)$ is
finite, there exists $\varepsilon > 0$ such that $\sigma^2(t) <
\varepsilon$ for any $t$. Therefore, $\sigma^2/ \langle\rho\rangle^2
\sim \varepsilon \,\textrm{e}^{2t/\tau}$ increases exponentially. This
observation hints about the general behavior of
$D_{\rho\rho}(t)$: the magnitude of $D_{\rho\rho}(t)$ should also
increase exponentially. We summarize these
observations by measuring $|D_{\rho\rho}/\langle \rho \rangle|^2$.
Fig.~\ref{fig:complete_normalized_corr} shows the striking differences
between symmetric and non-symmetric fluctuations. More importantly, for
non-symmetric fluctuations,    
\begin{equation}
  \label{eq:normalized2}
  D_{\rho\rho}(t)\approx -\alpha D_1\, \textrm{e}^{t/\tau} \langle  \rho(t)\rangle
\end{equation}
provides a convenient description for the normalized autocorrelation
function, with fitting parameters $D_1$ and $\tau $. We remark that
the exponential fitting in Eq.~(\ref{eq:normalized2})  
deviates from data values at the very beginning of the outbreak (see
Fig.~\ref{fig:complete_normalized_corr}), so there is still room for
improvements especially for more complex population structures.

\begin{figure}
  \includegraphics[width=0.95\columnwidth]{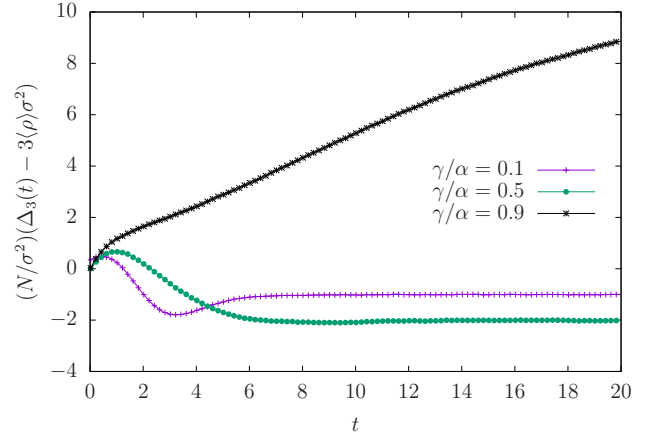}
  \caption{\label{fig:complete_delta3} Deviations from Gaussian
    behavior. Simulations are performed in the complete graph with
    $N=50$ agents, and $10^6$ samples. The quantity $\Delta_3 -
    \Delta_3^{\textrm{gauss}} = \sigma^2\kappa_3$ measures the
    deviation of the system compared to Gaussian fluctuations. Curves
    for $\gamma/\alpha = 0.1$ and $0.5$ imply $\kappa_3\sim
    o(1/N)$. This behavior is not observed for
    $\gamma/\alpha=0.9$. Error bars omitted. 
  }
\end{figure}

\textit{Gaussian fluctuations.} For large population sizes,
stochastic effects are entirely dominated by finite second moments and are well
represented by Gaussian fluctuations. Because they are distributed
according to a symmetric probability distribution function, their
coefficient of skewness vanishes, $\kappa_3(t)=0$. Noting that
$\kappa_3 =  (\Delta_3-  3\langle \rho   \rangle \sigma^2)/\sigma^3$,
we conclude that  $\Delta_3^{\textrm{gauss}} \approx  3 \langle
\rho(t)   \rangle \sigma^2(t)$ for Gaussian fluctuations. Indeed,
Fig.~\ref{fig:complete_delta3} shows the \textit{ansatz} is not too far
fetched since  $\Delta_3(t) - \Delta_3^{\textrm{gauss}}(t) \sim
o(\sigma^2/N)$ for ratios $\gamma/\alpha = 0.1$ and $0.5$, but not for
$\gamma/\alpha=0.9$, which according to Fig.~\ref{fig:complete_delta3}
are dominated by non-gaussian fluctuations.

Ignoring $o(1/N)$ corrections in Eq.~(\ref{eq:del-sigma}), we write
the following differential equations:
\begin{subequations}
  \label{eq:improved0}
  \begin{align}
    \label{eq:improved1}
  \frac{1}{\alpha}\frac{d}{dt}\ln  \langle\rho\rangle &=
  \rho_{\textrm{eq}} -\langle \rho\rangle - \frac{\sigma^2}{\langle
                                                        \rho\rangle},\\
    \label{eq:improved2}
  \frac{1}{2\alpha}\frac{d}{dt}\ln \sigma^2 &=  \rho_{\textrm{eq}}
  -2\langle \rho\rangle ,
\end{align}
\end{subequations}
valid under the assumption $N\rightarrow \infty$. We can draft an
immediate observation from Eq.~(\ref{eq:improved2}). As long as the
$\sigma^2(0) > 0$, uncertainties play a role in the SIS epidemic
model. Conversely, $\sigma^2(0) = 0$ implies $\sigma^2(t) = 0$ and
warrants the validity of Eq.~(\ref{eq:sis2}). In addition, we see that
the instantaneous Fano factor $\sigma^2(t)/\langle\rho(t)\rangle$ in
Eq.~(\ref{eq:improved1}) improves compartmental predictions if
$\sigma^2$ remains finite. However, it also emphasizes that
uncertainties rather than $\rho(t)$ drive the disease spreading in the
low-density regimes.

Despite the insights provided by Eqs.~(\ref{eq:improved1}) and
(\ref{eq:improved2}), there are still some remaining issues. The most
relevant issue deals with estimates for $\sigma^2(0)$ from
epidemiological data. This issue can be avoided entirely by combining
the system of differential equations into a single differential
equation (angular brackets dropped for simplicity):
\begin{equation}
  \label{eq:deldel-rho}
 \frac{d^2\rho}{d t^2} = 3\alpha\left( \rho_{\textrm{eq}} - 2 \rho
  \right)\left[\frac{d\rho}{dt} -\frac{2\alpha}{3}(\rho_{\textrm{eq}} - \rho)\rho \right].
\end{equation}
This equation shares the same steady state solution as
Eq.~(\ref{eq:sis}). The major difference occurs during the transient
regime: uncertainties introduced by Gaussian fluctuations slowdown
the system.

\textit{Conclusion.}
We investigate the effects of uncertainties to disease
spreading and their implications in the SIS epidemic model. We derive
stochastic equations for $\rho(t)$ and $\sigma^2(t)$ by introducing
surrogate dynamics  for symmetric and non-symmetric fluctuations.
Our findings reconcile the simplicity of canonical compartmental
equations  with the accuracy of agent-based simulations, thus creating
suitable tools for practitioners of Epidemiology and related
fields. 
At the core of this research, we demonstrate that uncertainty cannot be
neglected in the SIS epidemic model in finite populations, even when
the population is large and comprised of statistically equivalent agents.
For non-symmetric fluctuations, the normalized autocorrelation function
can be parametrized, providing again a closed system for the variables
$\langle \rho(t)\rangle$ and $\sigma^2(t)$. The special case of
Gaussian fluctuations provides additional simplifications from which
we derive a second-order differential equation for $\langle
\rho(t)\rangle$.
Finally, we stress that this research evaluates the impact of uncertainties
only for homogeneous populations. As a consequence, connections between
agents are described according to the complete graph. An intriguing
question that arises is whether the inherent uncertainties associated with
network metrics influence or enhance fluctuations in the disease
spreading. For instance, scale-free networks contain uncertainties
that scale with $N$, which in turn must contribute in the master
equation Eq.~(\ref{eq:fokker}).

\begin{acknowledgments}
  We are grateful for G Contesini comments during the manuscript
   preparation and subsequent discussions. GMN acknowledges Capes 88887.136416/2017-00,
   NDG thanks Capes for the financial support,
  and ASM acknowledges grants CNPq 307948/2014-5.   
\end{acknowledgments}


\begin{thebibliography}{13}%
\makeatletter
\providecommand \@ifxundefined [1]{%
 \@ifx{#1\undefined}
}%
\providecommand \@ifnum [1]{%
 \ifnum #1\expandafter \@firstoftwo
 \else \expandafter \@secondoftwo
 \fi
}%
\providecommand \@ifx [1]{%
 \ifx #1\expandafter \@firstoftwo
 \else \expandafter \@secondoftwo
 \fi
}%
\providecommand \natexlab [1]{#1}%
\providecommand \enquote  [1]{``#1''}%
\providecommand \bibnamefont  [1]{#1}%
\providecommand \bibfnamefont [1]{#1}%
\providecommand \citenamefont [1]{#1}%
\providecommand \href@noop [0]{\@secondoftwo}%
\providecommand \href [0]{\begingroup \@sanitize@url \@href}%
\providecommand \@href[1]{\@@startlink{#1}\@@href}%
\providecommand \@@href[1]{\endgroup#1\@@endlink}%
\providecommand \@sanitize@url [0]{\catcode `\\12\catcode `\$12\catcode
  `\&12\catcode `\#12\catcode `\^12\catcode `\_12\catcode `\%12\relax}%
\providecommand \@@startlink[1]{}%
\providecommand \@@endlink[0]{}%
\providecommand \url  [0]{\begingroup\@sanitize@url \@url }%
\providecommand \@url [1]{\endgroup\@href {#1}{\urlprefix }}%
\providecommand \urlprefix  [0]{URL }%
\providecommand \Eprint [0]{\href }%
\providecommand \doibase [0]{http://dx.doi.org/}%
\providecommand \selectlanguage [0]{\@gobble}%
\providecommand \bibinfo  [0]{\@secondoftwo}%
\providecommand \bibfield  [0]{\@secondoftwo}%
\providecommand \translation [1]{[#1]}%
\providecommand \BibitemOpen [0]{}%
\providecommand \bibitemStop [0]{}%
\providecommand \bibitemNoStop [0]{.\EOS\space}%
\providecommand \EOS [0]{\spacefactor3000\relax}%
\providecommand \BibitemShut  [1]{\csname bibitem#1\endcsname}%
\let\auto@bib@innerbib\@empty
\bibitem [{\citenamefont {Bonita}\ \emph {et~al.}(2006)\citenamefont {Bonita},
  \citenamefont {Beaglehole},\ and\ \citenamefont
  {Kjellstr{\"o}m}}]{bonita2006}%
  \BibitemOpen
  \bibfield  {author} {\bibinfo {author} {\bibfnamefont {R.}~\bibnamefont
  {Bonita}}, \bibinfo {author} {\bibfnamefont {R.}~\bibnamefont {Beaglehole}},
  \ and\ \bibinfo {author} {\bibfnamefont {T.}~\bibnamefont {Kjellstr{\"o}m}},\
  }\href@noop {} {\emph {\bibinfo {title} {Basic epidemiology}}}\ (\bibinfo
  {publisher} {World Health Organization},\ \bibinfo {year} {2006})\BibitemShut
  {NoStop}%
\bibitem [{\citenamefont {Albert}\ and\ \citenamefont
  {Barab\'asi}(2002)}]{albertRevModPhys2002}%
  \BibitemOpen
  \bibfield  {author} {\bibinfo {author} {\bibfnamefont {R.}~\bibnamefont
  {Albert}}\ and\ \bibinfo {author} {\bibfnamefont {A.-L.}\ \bibnamefont
  {Barab\'asi}},\ }\href {\doibase 10.1103/RevModPhys.74.47} {\bibfield
  {journal} {\bibinfo  {journal} {Rev. Mod. Phys.}\ }\textbf {\bibinfo {volume}
  {74}},\ \bibinfo {pages} {47} (\bibinfo {year} {2002})}\BibitemShut {NoStop}%
\bibitem [{\citenamefont {Pastor-Satorras}\ \emph {et~al.}(2015)\citenamefont
  {Pastor-Satorras}, \citenamefont {Castellano}, \citenamefont {Van~Mieghem},\
  and\ \citenamefont {Vespignani}}]{satorrasRevModPhys2015}%
  \BibitemOpen
  \bibfield  {author} {\bibinfo {author} {\bibfnamefont {R.}~\bibnamefont
  {Pastor-Satorras}}, \bibinfo {author} {\bibfnamefont {C.}~\bibnamefont
  {Castellano}}, \bibinfo {author} {\bibfnamefont {P.}~\bibnamefont
  {Van~Mieghem}}, \ and\ \bibinfo {author} {\bibfnamefont {A.}~\bibnamefont
  {Vespignani}},\ }\href {\doibase 10.1103/RevModPhys.87.925} {\bibfield
  {journal} {\bibinfo  {journal} {Rev. Mod. Phys.}\ }\textbf {\bibinfo {volume}
  {87}},\ \bibinfo {pages} {925} (\bibinfo {year} {2015})}\BibitemShut
  {NoStop}%
\bibitem [{\citenamefont {Eames}\ and\ \citenamefont
  {Keeling}(2002)}]{keelingPNAS2002}%
  \BibitemOpen
  \bibfield  {author} {\bibinfo {author} {\bibfnamefont {K.~T.~D.}\
  \bibnamefont {Eames}}\ and\ \bibinfo {author} {\bibfnamefont {M.~J.}\
  \bibnamefont {Keeling}},\ }\href {\doibase 10.1073/pnas.202244299} {\bibfield
   {journal} {\bibinfo  {journal} {Proc. Natl. Acad. Sci. USA}\ }\textbf
  {\bibinfo {volume} {99}},\ \bibinfo {pages} {13330} (\bibinfo {year}
  {2002})}\BibitemShut {NoStop}%
\bibitem [{\citenamefont {Keeling}\ and\ \citenamefont
  {Eames}(2005)}]{keelingJRSoc2005}%
  \BibitemOpen
  \bibfield  {author} {\bibinfo {author} {\bibfnamefont {M.}~\bibnamefont
  {Keeling}}\ and\ \bibinfo {author} {\bibfnamefont {K.}~\bibnamefont
  {Eames}},\ }\href {\doibase 10.1098/rsif.2005.0051} {\bibfield  {journal}
  {\bibinfo  {journal} {J. R. Soc. Interface}\ }\textbf {\bibinfo {volume}
  {2}},\ \bibinfo {pages} {295} (\bibinfo {year} {2005})}\BibitemShut {NoStop}%
\bibitem [{\citenamefont {Bansal}\ \emph {et~al.}(2007)\citenamefont {Bansal},
  \citenamefont {Grenfell},\ and\ \citenamefont {Meyers}}]{bansalJRSoc2007}%
  \BibitemOpen
  \bibfield  {author} {\bibinfo {author} {\bibfnamefont {S.}~\bibnamefont
  {Bansal}}, \bibinfo {author} {\bibfnamefont {B.~T.}\ \bibnamefont
  {Grenfell}}, \ and\ \bibinfo {author} {\bibfnamefont {L.~A.}\ \bibnamefont
  {Meyers}},\ }\href@noop {} {\bibfield  {journal} {\bibinfo  {journal} {J. R.
  Soc. Interface}\ }\textbf {\bibinfo {volume} {4}},\ \bibinfo {pages} {879}
  (\bibinfo {year} {2007})}\BibitemShut {NoStop}%
\bibitem [{\citenamefont {Kermack}\ and\ \citenamefont
  {McKendrick}(1927)}]{kermackProcRSocA1927}%
  \BibitemOpen
  \bibfield  {author} {\bibinfo {author} {\bibfnamefont {W.~O.}\ \bibnamefont
  {Kermack}}\ and\ \bibinfo {author} {\bibfnamefont {A.~G.}\ \bibnamefont
  {McKendrick}},\ }\href {\doibase 10.1098/rspa.1927.0118} {\bibfield
  {journal} {\bibinfo  {journal} {Proc. R. Soc. A}\ }\textbf {\bibinfo {volume}
  {115}},\ \bibinfo {pages} {700} (\bibinfo {year} {1927})}\BibitemShut
  {NoStop}%
\bibitem [{\citenamefont {Pastor-Satorras}\ and\ \citenamefont
  {Vespignani}(2001{\natexlab{a}})}]{satorrasPhysRevLett2001}%
  \BibitemOpen
  \bibfield  {author} {\bibinfo {author} {\bibfnamefont {R.}~\bibnamefont
  {Pastor-Satorras}}\ and\ \bibinfo {author} {\bibfnamefont {A.}~\bibnamefont
  {Vespignani}},\ }\href {\doibase 10.1103/PhysRevLett.86.3200} {\bibfield
  {journal} {\bibinfo  {journal} {Phys. Rev. Lett.}\ }\textbf {\bibinfo
  {volume} {86}},\ \bibinfo {pages} {3200} (\bibinfo {year}
  {2001}{\natexlab{a}})}\BibitemShut {NoStop}%
\bibitem [{\citenamefont {Pastor-Satorras}\ and\ \citenamefont
  {Vespignani}(2001{\natexlab{b}})}]{satorrasPhysRevE2001}%
  \BibitemOpen
  \bibfield  {author} {\bibinfo {author} {\bibfnamefont {R.}~\bibnamefont
  {Pastor-Satorras}}\ and\ \bibinfo {author} {\bibfnamefont {A.}~\bibnamefont
  {Vespignani}},\ }\href {\doibase 10.1103/PhysRevE.63.066117} {\bibfield
  {journal} {\bibinfo  {journal} {Phys. Rev. E}\ }\textbf {\bibinfo {volume}
  {63}},\ \bibinfo {pages} {066117} (\bibinfo {year}
  {2001}{\natexlab{b}})}\BibitemShut {NoStop}%
\bibitem [{\citenamefont {Chen}\ \emph {et~al.}(2017)\citenamefont {Chen},
  \citenamefont {Ghanbarnejad},\ and\ \citenamefont
  {Brockmann}}]{brockmannNJPhys2017}%
  \BibitemOpen
  \bibfield  {author} {\bibinfo {author} {\bibfnamefont {L.}~\bibnamefont
  {Chen}}, \bibinfo {author} {\bibfnamefont {F.}~\bibnamefont {Ghanbarnejad}},
  \ and\ \bibinfo {author} {\bibfnamefont {D.}~\bibnamefont {Brockmann}},\
  }\href {http://stacks.iop.org/1367-2630/19/i=10/a=103041} {\bibfield
  {journal} {\bibinfo  {journal} {New Journal of Physics}\ }\textbf {\bibinfo
  {volume} {19}},\ \bibinfo {pages} {103041} (\bibinfo {year}
  {2017})}\BibitemShut {NoStop}%
\bibitem [{\citenamefont {Heesterbeek}\ \emph {et~al.}(2015)\citenamefont
  {Heesterbeek}, \citenamefont {Anderson}, \citenamefont {Andreasen},
  \citenamefont {Bansal}, \citenamefont {De~Angelis}, \citenamefont {Dye},
  \citenamefont {Eames}, \citenamefont {Edmunds}, \citenamefont {Frost},
  \citenamefont {Funk}, \citenamefont {Hollingsworth}, \citenamefont {House},
  \citenamefont {Isham}, \citenamefont {Klepac}, \citenamefont {Lessler},
  \citenamefont {Lloyd-Smith}, \citenamefont {Metcalf}, \citenamefont
  {Mollison}, \citenamefont {Pellis}, \citenamefont {Pulliam}, \citenamefont
  {Roberts},\ and\ \citenamefont {Viboud}}]{heesterbeekScience2015}%
  \BibitemOpen
  \bibfield  {author} {\bibinfo {author} {\bibfnamefont {H.}~\bibnamefont
  {Heesterbeek}}, \bibinfo {author} {\bibfnamefont {R.~M.}\ \bibnamefont
  {Anderson}}, \bibinfo {author} {\bibfnamefont {V.}~\bibnamefont {Andreasen}},
  \bibinfo {author} {\bibfnamefont {S.}~\bibnamefont {Bansal}}, \bibinfo
  {author} {\bibfnamefont {D.}~\bibnamefont {De~Angelis}}, \bibinfo {author}
  {\bibfnamefont {C.}~\bibnamefont {Dye}}, \bibinfo {author} {\bibfnamefont
  {K.~T.~D.}\ \bibnamefont {Eames}}, \bibinfo {author} {\bibfnamefont {W.~J.}\
  \bibnamefont {Edmunds}}, \bibinfo {author} {\bibfnamefont {S.~D.~W.}\
  \bibnamefont {Frost}}, \bibinfo {author} {\bibfnamefont {S.}~\bibnamefont
  {Funk}}, \bibinfo {author} {\bibfnamefont {T.~D.}\ \bibnamefont
  {Hollingsworth}}, \bibinfo {author} {\bibfnamefont {T.}~\bibnamefont
  {House}}, \bibinfo {author} {\bibfnamefont {V.}~\bibnamefont {Isham}},
  \bibinfo {author} {\bibfnamefont {P.}~\bibnamefont {Klepac}}, \bibinfo
  {author} {\bibfnamefont {J.}~\bibnamefont {Lessler}}, \bibinfo {author}
  {\bibfnamefont {J.~O.}\ \bibnamefont {Lloyd-Smith}}, \bibinfo {author}
  {\bibfnamefont {C.~J.~E.}\ \bibnamefont {Metcalf}}, \bibinfo {author}
  {\bibfnamefont {D.}~\bibnamefont {Mollison}}, \bibinfo {author}
  {\bibfnamefont {L.}~\bibnamefont {Pellis}}, \bibinfo {author} {\bibfnamefont
  {J.~R.~C.}\ \bibnamefont {Pulliam}}, \bibinfo {author} {\bibfnamefont
  {M.~G.}\ \bibnamefont {Roberts}}, \ and\ \bibinfo {author} {\bibfnamefont
  {C.}~\bibnamefont {Viboud}},\ }\href@noop {} {\bibfield  {journal} {\bibinfo
  {journal} {Science}\ }\textbf {\bibinfo {volume} {347}} (\bibinfo {year}
  {2015})}\BibitemShut {NoStop}%
\bibitem [{\citenamefont {Nakamura}\ \emph {et~al.}(2017)\citenamefont
  {Nakamura}, \citenamefont {Monteiro}, \citenamefont {Cardoso},\ and\
  \citenamefont {Martinez}}]{nakamuraSciRep2017}%
  \BibitemOpen
  \bibfield  {author} {\bibinfo {author} {\bibfnamefont {G.~M.}\ \bibnamefont
  {Nakamura}}, \bibinfo {author} {\bibfnamefont {A.~C.~P.}\ \bibnamefont
  {Monteiro}}, \bibinfo {author} {\bibfnamefont {G.~C.}\ \bibnamefont
  {Cardoso}}, \ and\ \bibinfo {author} {\bibfnamefont {A.~S.}\ \bibnamefont
  {Martinez}},\ }\href {\doibase 10.1038/srep40885} {\bibfield  {journal}
  {\bibinfo  {journal} {Scientific Reports}\ }\textbf {\bibinfo {volume} {7}},\
  \bibinfo {pages} {40885} (\bibinfo {year} {2017})}\BibitemShut {NoStop}%
\bibitem [{\citenamefont {Blanter}\ and\ \citenamefont
  {Büttiker}(2000)}]{blanterPhysRep2000}%
  \BibitemOpen
  \bibfield  {author} {\bibinfo {author} {\bibfnamefont {Y.}~\bibnamefont
  {Blanter}}\ and\ \bibinfo {author} {\bibfnamefont {M.}~\bibnamefont
  {Büttiker}},\ }\href {\doibase
  https://doi.org/10.1016/S0370-1573(99)00123-4} {\bibfield  {journal}
  {\bibinfo  {journal} {Physics Reports}\ }\textbf {\bibinfo {volume} {336}},\
  \bibinfo {pages} {1 } (\bibinfo {year} {2000})}\BibitemShut {NoStop}%
\end{thebibliography}

%

\end{document}